\documentclass[11pt]{article}
\usepackage{moriond,epsfig}

\usepackage{latexsym,amssymb,amsfonts,amsmath}

\def\NCA{{\em Nuovo Cimento}}

\def\NPA{{\em Nucl. Phys.} A}
\def\NPB{{\em Nucl. Phys.} B}
\def\PLB{{\em Phys. Lett.} B}
\def\PRL{{\em Phys. Rev. Lett.}}
\def\PRD{{\em Phys. Rev.} D}
\def\ZPC{{\em Z. Phys.} C}

\def\be{\begin{equation}}
\def\ee{\end{equation}}
\def\bea{\begin{eqnarray}}
\def\eea{\end{eqnarray}}

\def\nin{\noindent}
\def\beq{\begin{equation}}
\def\eeq{\end{equation}}
\def\bea{\begin{eqnarray}}
\def\eea{\end{eqnarray}}

\newcommand{\virg}[1]{``#1''}
\newcommand{\trpl}[1]{\vec{#1}_\perp}

\newcommand{\re}{\text{Re}}
\newcommand{\im}{\text{Im}}
\newcommand{\tr}{\text{Tr}}
\newcommand{\lla}{\langle\langle}
\newcommand{\rra}{\rangle\rangle}
\begin{document}
\vspace*{4cm}
\title{ASYMPTOTIC HIGH-ENERGY BEHAVIOR OF HADRONIC TOTAL CROSS SECTIONS
FROM LATTICE QCD}




\author{\underline{ENRICO MEGGIOLARO}$^a$, MATTEO GIORDANO$^b$,
NICCOL\`O MORETTI$^c$}
\address{$^a$Dipartimento di Fisica, Universit\`a di Pisa, and
INFN, Sezione di Pisa,\\ Largo Pontecorvo 3, I-56127 Pisa, Italy\\
$^b$Institute of Nuclear Research of the Hungarian Academy of Sciences
(ATOMKI),\\ Bem t\'er 18/c, H-4026 Debrecen, Hungary\\
$^c$Institut f\"ur Theoretische Physik, Universit\"at Z\"urich,
8057 Z\"urich, Switzerland}

\maketitle\abstracts{
We will show how a {\it universal} and {\it Froissart-like} (i.e., of the kind
$B \log^2 s$) hadron-hadron total cross section can emerge in QCD asymptotically
at high energy, finding indications for this behavior from the lattice.
The functional integral approach provides the \virg{natural} setting for
achieving this result, since it encodes the energy dependence of hadronic
scattering amplitudes in a single \textit{elementary} object, i.e., a proper
correlation function of two Wilson loops.}

\section{Introduction}
\nin
Present-day experimental observations (up to a center-of-mass total energy
$\sqrt{s} = 7$ TeV, reached at the LHC $pp$ collider~\cite{LHC})
seem to support the following asymptotic high-energy behavior of hadronic
total cross sections:
$\sigma_{\rm tot}^{{}_{(hh)}} (s) \sim B \log^2 s$, with a {\it universal}
(i.e., {\it not} depending on the particular hadrons
involved) coefficient $B \simeq 0.3$ mb.~\cite{Blogs} This behavior is
consistent with the well-known {\it Froissart-Lukaszuk-Martin} (FLM)
{\it theorem}~\cite{FLM}, according to which, for $s \to \infty$,
$\sigma^{{}_{(hh)}}_{\rm tot}(s) \le ({\pi}/{m_\pi^2}) \log^2 (
{{s}/{s_0}} )$, where $m_\pi$ is the pion mass and $s_0$ is an
unspecified squared mass scale. As we believe QCD to be the
fundamental theory of strong interactions, we also expect that it
correctly predicts from first principles the behavior of hadronic
total cross sections. However, in spite of all the efforts, a
satisfactory solution to this problem is still lacking.
(For some theoretical supports to the universality of $B$,
see Ref.~\cite{DGN} and references therein.)

This problem is part of the more general problem of high-energy
elastic scattering at low transferred momentum, the so-called {\it soft
high-energy scattering}. As {\it soft} high-energy processes possess
two different energy scales, the total center-of-mass energy squared 
$s$ and the transferred momentum squared $t$, smaller than the typical 
energy scale of strong interactions ($|t| \lesssim 1~ {\rm GeV}^2 \ll
s$), we cannot fully rely on perturbation theory (PT). A nonperturbative (NP)
functional-integral approach in the framework of QCD has been
proposed in Ref.~\cite{Nachtmann91} and further developed in Ref.~\cite{DFK}
In this approach, for example, the elastic scattering amplitude
${\cal M}_{(hh)}$ of two {\it mesons}, of the same mass $m$ for
simplicity, can be reconstructed from the scattering amplitude
${\cal M}_{(dd)}$ of two dipoles of fixed transverse sizes
$\vec{r}_{1,2\perp}$, and fixed longitudinal-momentum fractions
$f_{1,2}$ of the quarks in the two dipoles, after folding with squared
wave functions $\rho_{1,2}=|\psi_{1,2}|^2$ describing the interacting
hadrons,~\cite{DFK}
\begin{equation}
{\cal M}_{(hh)}(s,t) = \textstyle\int
d^2\nu~\rho_1(\nu_1)\rho_2(\nu_2){\cal M}_{(dd)} (s,t;\nu_1,\nu_2)
\equiv \lla {\cal M}_{(dd)} (s,t;1,2) \rra ,
\label{scatt-hadron}
\end{equation}
where $\nu_i\!=\!(\vec{r}_{i\perp},f_i)$ denotes collectively the
dipole variables, $d^2\nu=d\nu_1d\nu_2$, $\int d\nu_i = \int
d^2\vec{r}_{i\perp}\int_0^1 df_i$, and $\int d\nu_i~\rho_i(\nu_i)=1$.
In turn, the dipole-dipole ({\it dd\/}) scattering amplitude is obtained from
the (properly normalized) correlation function (CF) of two Wilson loops (WL)
in the fundamental representation, defined in
Minkowski spacetime, running along the paths made up of the quark and
antiquark classical straight-line trajectories, and thus forming a
hyperbolic angle $\chi \simeq \log(s/m^2)$ in the longitudinal plane. 
The paths are cut at proper times $\pm T$ as an infrared
regularization, and closed by straight-line ``links'' in the
transverse plane, in order to ensure gauge invariance; eventually,
$T\to\infty$.
It has been shown in Refs.~\cite{Meggiolaro1997-2002,Meggiolaro2005,GM2006-2009}
that the relevant Minkowskian CF ${\cal G}_M(\chi;T;\vec{z}_\perp;\nu_1,\nu_2)$
($\vec{z}_\perp$ being the {\it impact parameter}, i.e., the transverse
separation between the two dipoles) can be reconstructed, by means of
{\it analytic continuation}, from the Euclidean CF of two Euclidean WL,
${\cal G}_E(\theta;T;\vec{z}_\perp;\nu_1,\nu_2) \!\equiv\! 
\langle {\cal W}^{{}_{(T)}}_1 {\cal W}^{{}_{(T)}}_2\rangle/(\langle
{\cal W}^{{}_{(T)}}_1 \rangle \langle {\cal W}^{{}_{(T)}}_2 \rangle ) - 1$,
where $\langle\ldots\rangle$ is the average in the sense of the
Euclidean QCD functional integral. The Euclidean WL 
${\cal W}^{{}_{(T)}}_{1,2}\!=\! N_c^{-1}\tr\{T\!\exp [-ig\oint_{{\cal C}_{1,2}}
\!{A}_{\mu}({x}) d{x}_{\mu}]\}$
are calculated on the following quark $[q]$-antiquark $[\bar{q}]$
straight-line paths, ${\cal C}_i : 
{X}_i^{{}_{q[\bar{q}]}}(\tau) = {z}_i + \frac{{p}_{i}}{m} \tau +
f^{{}_{q[\bar{q}]}}_i {r}_{i}$, 
with $\tau\in [-T,T]$, and closed by straight-line paths in the
transverse plane at $\tau=\pm T$. Here
${p}_{1,2}={m}(\pm\sin\frac{\theta}{2}, \vec{0}_{\perp}, \cos\frac{\theta}{2})$,
$\theta$ being the angle formed by the two Euclidean trajectories
(i.e., $p_1 \cdot p_2 = m^2 \cos\theta$),
${r}_{i} = (0,\vec{r}_{i\perp},0)$, ${z}_i =
\delta_{i1}(0,\vec{z}_{\perp},0)$ and $f_i^{{}_{q}} \equiv 1-f_i$,
$f_i^{{}_{\bar{q}}} \equiv -f_i$. 
We define also the CFs with the infrared cutoff removed as ${\cal C}_{E,M}
\equiv\lim_{T\to\infty}{\cal G}_{E,M}$. The {\it dd}
scattering amplitude is then obtained from ${\cal C}_E(\theta;\ldots)$
[with $\theta\in(0,\pi)$] by means of analytic continuation as ($t =
-|\vec{q}_\perp|^2$) 
\begin{align}
&{\cal M}_{(dd)} (s,t;\nu_1,\nu_2) 
\!\equiv\! -i\,2s \textstyle\int d^2 \vec{z}_\perp
e^{i \vec{q}_\perp \cdot \vec{z}_\perp}
{\cal C}_M(\chi ; \vec{z}_\perp;\nu_1,\nu_2) 
\nonumber \\ \label{scatt-loop}
&= -i\,2s \textstyle\int d^2 \vec{z}_\perp
e^{i \vec{q}_\perp \cdot \vec{z}_\perp}
{\cal C}_E(\theta\to -i\chi ; \vec{z}_\perp;\nu_1,\nu_2) .
\end{align}
In Refs.~\cite{GM2008,GM2010} the CF ${\cal C}_E$ were calculated
in {\it quenched} QCD by Monte Carlo simulations in
\textit{Lattice Gauge Theory} (LGT), at lattice spacing
$a(\beta=6)\simeq 0.1\,{\rm fm}$, on a $16^4$ hypercubic lattice,
using loops of transverse size $a$ at angles $\cot \theta\! =\! 0,
\pm \frac{1}{2}, \pm 1,\pm 2$ and transverse distances
$d \equiv |\vec{z}_\perp|/a = 0,1,2$.
The longitudinal-momentum fractions were set to
$f_{1,2}=\frac{1}{2}$ without loss of generality~\cite{GM2010}
and different configurations in the transverse plane were studied,
including the one relevant to meson-meson scattering, that is the
{\it average} over the transverse orientations (``{\it ave}'').

Numerical simulations of LGT provide (within the errors) the true QCD
expectation for ${\cal C}_E$; approximate analytical calculations of
${\cal C}_E$ have then to be compared with the lattice data, in order
to test the goodness of the approximations involved. ${\cal C}_E$ has
been evaluated in the \textit{Stochastic Vacuum Model} (SVM),~\cite{LLCM2}
${\cal C}^{\rm{}_{(SVM)}}_E\!=\! \textstyle\frac{2}{3}e^{-\frac{1}{3}
K_{\rm S}\cot\theta} + \frac{1}{3} e^{\frac{2}{3} K_{\rm S}\cot\theta} - 1$,
in PT,~\cite{Meggiolaro2005,LLCM2,BB}
${\cal C}_E^{\rm {}_{(PT)}}\! =\! K_{\rm p} \cot^2\theta$,
in the \textit{Instanton Liquid Model} (ILM),~\cite{GM2010,ILM}
${\cal C}^{\rm {}_{(ILM)}}_E\!=\! \frac{K_{\rm I}}{\sin\theta}$,
and, using the AdS/CFT correspondence, for planar, strongly coupled
${\cal N}=4$ SYM at large $|\trpl{z}|$,~\cite{JP}
${\cal C}^{\rm {}_{(AdS/CFT)}}_E = e^{\frac{K_1}{\sin\theta} +
K_2\cot\theta + K_3\cos\theta\cot\theta}-1$.
The coefficients $K_i = K_i(\trpl{z};\nu_1,\nu_2)$ are functions of
$\trpl{z}$ and of the dipole variables $\vec{r}_{i\perp}, f_i$. The
comparison of the lattice data with these analytical calculations,
performed in Ref.~\cite{GM2008} by fitting the lattice data with the
corresponding functional form, is not fully satisfactory, even though
largely improved best fits have been obtained by combining the ILM and
PT expressions into the expression ${\cal C}^{\rm {}_{(ILMp)}}_E =
\frac{K_{\rm Ip1}}{\sin\theta} + K_{\rm Ip2}\cot^2\theta $. 
Regarding the energy dependence of total cross
sections, the above analytical models are absolutely unsatisfactory,
as they do not lead to {\it Froissart-like} total cross sections at
high energy, as experimental data seem to suggest. Infact, the SVM,
PT, ILM and ILMp parameterizations lead to asymptotically constant
$\sigma_{\rm tot}^{{}_{(hh)}}$, while the AdS/CFT result leads to
power-like $\sigma_{\rm tot}^{{}_{(hh)}}$.~\cite{GP2010}

\section{How a Froissart-like total cross section can be obtained} 

\nin
One is thus motivated to look for new parameterizations of the CF that: 
i) fit well the data;
ii) satisfy the unitarity condition after analytic continuation; and 
iii) lead to total cross sections rising as $B \log^2 s$ in the
high-energy limit.~\cite{GMM2012}
Regarding unitarity, from~\eqref{scatt-hadron} and \eqref{scatt-loop}
one recognizes \newpage
\nin
that the quantity $A(s,|\trpl{z}|)\equiv
\lla\mathcal{C}_M(\chi;\trpl{z};1,2)\rra$ is the scattering
amplitude in impact-parameter space, which must satisfy the
{\it unitarity constraint} $|A+1|\leq 1$.~\cite{unitarity} Since
$\int d\nu_i~\rho_i(\nu_i)=1$, this is the case if the following
\textit{sufficient} condition is satisfied:
$|\mathcal{C}_M(\chi;\trpl{z};\nu_1,\nu_2)+1| \leq 1$,
$\forall \trpl{z},~\nu_1,~\nu_2$.
The conditions above constrain rather strongly the possible
parameterizations.
We shall \textit{assume} that the Euclidean CF can be
written as $\mathcal{C}_E=\exp K_E-1$, where
$K_E=K_E(\theta;\trpl{z};\nu_1,\nu_2)$ is a \textit{real} function
(since $\mathcal{C}_E$ is \textit{real}~\cite{GM2008}). This assumption
is rather well justified: in the large-$N_c$ expansion, $\mathcal{C}_E
\sim\mathcal{O}(1/N_c^2)$, so that $\mathcal{C}_E+1\geq0$ is certainly
satisfied for large $N_c$; all the known analytical models satisfy it; 
the lattice data of Refs.~\cite{GM2008,GM2010} confirm it. 
The Minkowskian CF is then obtained after analytic continuation:
$\mathcal{C}_M=\exp K_M-1$, with $K_M(\chi;\ldots) = K_E(\theta \to\! 
-i\chi;\ldots)$. At large $\chi$, $\mathcal{C}_M$ is expected to obey
the above-mentioned unitarity condition, which in this case
reduces to $\re K_M\leq0\,\,\, \forall \trpl{z}, \nu_1,\nu_2$.

For a {\it confining} theory like QCD, $\mathcal{C}_E$ is expected to
decay exponentially as $\mathcal{C}_E\sim(\sum)\,e^{-\mu |\trpl{z}|}$
at large $|\trpl{z}|$, with mass scales $\mu$ related to the masses of
particles (including, possibly, also {\it glueballs}~\cite{glueballs})
exchanged between the two WL.
Therefore, one also expects a similar large-$|\trpl{z}|$ behavior for $K_E$,
i.e., $K_E \sim (\sum)\,e^{-\mu |\trpl{z}|}$.
(Instead, for a non-confining, let's say {\it conformal}, field theory,
different behaviors like powers of $1/|\trpl{z}|$ are typical.~\cite{JP,GP2010})

Let us now assume that the leading term of the Minkowskian CF for
$\chi\!\to\!+\infty$ is of the form $\mathcal{C}_M \sim \exp\big(i\,
\beta\,f (\chi)\,e^{-\mu |\trpl{z}|}\big)-1$ (i.e.,
$K_M \sim i\,\beta\,f (\chi)\,e^{-\mu |\trpl{z}|}$)
where $\beta\!=\!\beta(\nu_1,\nu_2)$ is a
function of the dipole variables and $f(\chi)$ is a \textit{real}
function such that $f(\chi)\to\!+\infty$ for $\chi\to\! +\infty$.
In this case, the unitarity condition is equivalent 
(for large $\chi$) to {$\im\beta\!\geq\!0$}.
By virtue of the {\it optical theorem},
$\sigma_{\rm tot}^{{}_{(hh)}} (s)\!\sim \! s^{-1} {\rm Im} {\cal M}_{(hh)}
(s, t\!=\!0)$, we find $\sigma^{{}_{(hh)}}_{\text{tot}} \sim
{4\pi}{\mu^{-2}}\textstyle\lla \textstyle\frac{1}{2}\log^2
f(\chi)+\log f(\chi) (\log|\beta| + \gamma)+\dots\rra$. If one takes
$f(\chi)= \chi^p e^{n\chi}$, the resulting asymptotic behavior of
$\sigma^{{}_{(hh)}}_{\text{tot}} $ is
[recall $\chi\simeq\log(s/m^2)$]~\cite{GMM2012}
\begin{equation}\label{sigmatotlead}
\vspace{-1.48pt}
\sigma^{(hh)}_{\text{tot}} \sim B \log^2 s,
\qquad \text{with:}\qquad B = \textstyle\frac{2\pi n^2}{\mu^2}.
\vspace{-1.48pt}
\end{equation}
We want to emphasize that the above result is \textit{universal}, depending
only on the mass scale $\mu$, which sets the large-$|\trpl{z}|$ dependence of
the leading term of the CF, since the integration
over the dipole variables does {\it not} affect the leading term. 
The \textit{universal} coefficient $B$ is not affected by the masses of
the scattering particles: for mesons of masses $m_{1,2}$, the rapidity
becomes $\chi\sim\log(\frac{s}{m_1 m_2})$, which simply corresponds to
a change of the energy scale implicitly contained in~\eqref{sigmatotlead}.
This result can also be extended to the case in which $K_M$ is, for large
$\chi$, the sum of different terms, each behaving like the one discussed
above, but with different values of $n$ and $\mu$, i.e.,
$K_M \sim i\sum_k\beta_k\chi^{p_k}e^{n_k\chi}e^{-\mu_k |\trpl{z}|}$:
the resulting $B$ comes out to be determined by the maximum value of the
ratio $n/\mu$ among these terms, i.e.,
$B = 2\pi\max_k(\frac{n_k}{\mu_k})^2$.
From a physical point of view, one expects that particles with {\it mass} $M$
and {\it spin} $J$, exchanged between the two WL, contribute with $\mu=M$
and $n=J-1$: in fact, in this case the factor $e^{n\chi}$ reduces to the
well-known factor $s^{J-1}$, expected for the contribution of an exchanged
particle of spin $J$.~\cite{GMM2013}

\section{New analysis of the lattice data}

\nin
In Ref.~\cite{GMM2012} we have found three parameterizations
${\cal C}_E^{{}_{(i)}}=\exp{K_E^{{}_{(i)}}}-1$, $i=1,2,3$
(among the many that we have analyzed),
that satisfy the criteria i)--iii) listed above.
We have focused our analysis on the {\it averaged} CF $\mathcal{C}^{ave}$,
that is \virg{closer} to the hadronic scattering matrix ${\cal M}_{(hh)}$.

The first two parameterizations, 
$K_E^{{}_{(1)}}=\frac{K_1}{\sin\theta}+K_2 \cot^2\theta +
K_3 \cos\theta\cot\theta$ and
$K_E^{{}_{(2)}} = \frac{K_1}{\sin\theta} + K_2 (\frac{\pi}{2}-\theta)
\cot\theta + K_3 \cos\theta\cot\theta$,
are essentially two proper modifications of the AdS/CFT result.
The third parameterization is, instead, $K_E^{{}_{(3)}} = \frac{K_1}{\sin\theta}
+ K_2(\frac{\pi}{2}-\theta)^3 \cos\theta$: while the first term is
\virg{familiar}, the second one is not present in the known analytical
models, but it is a fact that the resulting best fit is extremely good.
In the three cases, the unitarity condition $\re K_M^{{}_{(i)}}\le 0$
is satisfied if $K_2\geq0$: this is actually the case for our best
fits (within the errors). The leading term after analytic continuation
(the third term in the first two parameterizations $K^{{}_{(1)}}$ and
$K^{{}_{(2)}}$ and the second term in the parameterization $K^{{}_{(3)}}$)
is of the form $\chi^p e^{\chi}$ for large $\chi$, which, according to the
previous discussion, should correspond to an exchanged particle of spin $J=2$
(being $n=1$), and, according
to~\eqref{sigmatotlead}, leads to $\sigma^{{}_{(hh)}}_{\text{tot}}
\sim B \log^2 s$. The value of $B=2\pi/\mu^2$, obtained through a fit
of the coefficient of the leading term with an exponential function
$\sim e^{-\mu |\trpl{z}|} $ over the available distances, is found to
be compatible with the experimental result $B_{\rm exp} \simeq 0.3$ mb
(within the large errors) in all the three cases
(see Table~\ref{tab:lambdavac}). However, this
must be taken only as an estimate, as lattice data are available only
for small $|\trpl{z}|$.
Since $\mu_{\rm exp} = \sqrt{2\pi/B_{\rm exp}} \simeq 2.85$ GeV is close to
the value of the mass for the {\it glueball} $2^{++}$,~\cite{glueballs}
one could be tempted to conclude, on the basis of the results that we have
found, that $B$ is determined by the mass of this {\it glueball}.
But this is maybe still a premature conclusion: work is in progress
along this direction.~\cite{GMM2013}


\begin{table}
\centering
\caption{Mass-scale $\mu$, \virg{decay length} $\lambda=1/\mu$ and
the coefficient $B=2\pi/\mu^2$ obtained with our parameterizations.}
\vspace{0.4cm}
\begin{tabular}{|l|c|c|c|}
\hline
& $\mu$ (GeV) & $\lambda=\frac{1}{\mu}$ (fm) & $B=\frac{2\pi}{\mu^2}$ (mb) \\
\hline
Corr 1 & $4.64(2.38)$ & $0.042^{+0.045}_{-0.014}$ & $0.113^{+0.364}_{-0.037}$ \\
Corr 2 & $3.79(1.46)$ & $0.052^{+0.032}_{-0.014}$ & $0.170^{+0.277}_{-0.081}$ \\
Corr 3 & $3.18(98)$ & $0.062^{+0.028}_{-0.015}$ & $0.245^{+0.263}_{-0.100}$ \\
\hline
\end{tabular}
\label{tab:lambdavac}
\end{table}

\end{document}